# Anatomy of Demagnetizing and Exchange Fields in Magnetic Nano-dots Influenced by 3D Shape Modifications

T. Blachowicz

*Institute of Physics, Silesian University of Technology*
*Krzywoustego 2 str., 44100 Gliwice, Poland*
*e-mail: tomasz.blachowicz@polsl.pl*

A. Ehrmann

*Faculty for Textile and Clothing Technology, Niederrhein University of Applied Sciences*
*41065 Mönchengladbach, Germany*

(Dated: July 19, 2012)

Hysteresis loops of 3D ferromagnetic permalloy nano-half-balls (dots) with 100 nm base diameter have been examined by means of LLG micromagnetic simulations and finite element methods. Tests were carried out with two orthogonal directions of the externally applied field at 10 kA/(m·ns) field sweeping speed. The comparison of samples with different 3D modifications at the sub-10nm scale, accessible by nowadays lithographic techniques, enables conclusions about different mechanisms of competition between demagnetizing and exchange fields. Design paradigms provided here can find possible applications in magneto-electronic devices.

## I. INTRODUCTION

Analysis of the microscopic structure of magnetic states including exchange and demagnetization contributions plays an important role in understanding magnetic states in low dimensional objects. Observed in nano-dots and rings, flux-closure states, vortexes, and onion states were extensively investigated in many experimental works.[1-5] Additionally, methods allowing for magnetization dynamics control for different disk shapes were realized in theory and experiment.[6-11] From those efforts emerges the need to deeply understand effects resulting from broken spatial symmetry of magnetic objects and subsequent changes of the reversal dynamics and the magnetization states at remanence.

Due to their importance in sensing and data storage applications,[12-14] nanostructured ferromagnetic (FM) elements have recently been examined theoretically and experimentally in several studies.[15-17] Their magnetic properties can differ severely from those of bulk or thin film samples and depend on dimensions and shape of the nanoparticles,[18] allowing for desired properties to be created by tailoring the respectively formed nano-magnets. While most of the recent engineering efforts concentrated on 2D nano-devices, like dots,[19] rings,[4] wires,[20] etc., the first experiments with thin magnetic films on non-magnetic spheres with diameters between 20 and 1000 nm underline the importance of probing 3D nano-magnets.[21-22] The 3D nano-objects can principally be structured with today's techniques inspired by biologically self-assembled processes using proper (smart) components.[23]

It is worth to mention the nano-imprint technology, one of the most challenging techniques, for serial and cheap production of magnetic nano-objects.[24-26] The technology introduced in mid 1990s can reproduce sub-10nm local shape-features with very good precision.[27-28] The lithographic methods of serial objects, in general, can find applications in patterned magnetic recording media.[29-30]

The idea of the study presented here is based on the effect that for small magnetic elements the shape modifications – due to dipolar interactions – play a crucial role in magnetization dynamics, however, due to reduced dimensionality, the demagnetizing fields can compete with exchange fields leading to specific magnetic behaviours; such as regular and irregular oscillations, multi-frequency oscillations, rapid transient states, and quasi-static states at remanence or saturation.

Thus, we theoretically examine 3D ferromagnetic half-balls made of commonly applied permalloy (Py) with different shape modifications in order to support development of new 3D performance paradigms. In our samples the circular base of diameter 100 nm is located in the x-y-plane. In some of the half-balls, a perpendicular hole of diameter 50 nm has been cut in order to extract the core region of a possible magnetic vortex state which is known to have a comparable diameter.[31] In that way the samples represent all basic shape features to be met in 3D-fabricated nanomagnets: planes, curved surfaces, straight and bent edges, as well as intentionally introduced cuts and holes. In order to narrow the analysis to magnetostatic vs. exchange-based competition, we analyzed permalloy samples excluding magnetocrystalline anisotropy. It should be mentioned, however, that for nano-sized polycrystalline elements, the finite number of crystallites can be a source of additional, local magnetocrystalline anisotropies. Nevertheless, in our considerations these contributions were also excluded. All deviations from an ideal half-ball, created in different ways, can be seen in Fig. 1.

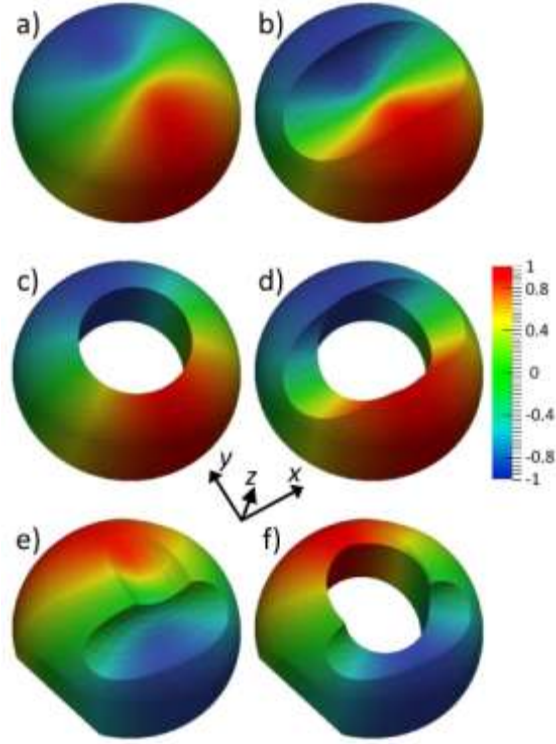

FIG. 1: The samples under simulation: (a), (b) and (e) are solid half-balls, while (c) and (d) half-balls possess a 50 nm diameter hole, respectively. The height of the half-ball (a) equals 50 nm. The cylindrical cuts in the x-y plane (b, d) have 50 nm diameter, and the maximum depth of a cut, seen along the half-ball z-axis, equals 25 nm with respect to the top point of the bulk half-ball (a). Samples (e) and (f) possess arbitrary, asymmetrical cuts, and the elliptical hole in (f) case is not parallel to the z-axis. The surface x-components of magnetization, at zero-valued externally applied magnetic field, are colour-coded.

Here we report results of hysteretic behaviour of the half-balls using micromagnetic simulations,[32] allowing for an insight into the magnetic properties, basing on the Landau-Lifshitz-Gilbert (LLG) equation of motion

$$\frac{\partial \vec{J}}{\partial t} = -\frac{\gamma}{1+\alpha^2}\vec{J}\times\vec{H}_{eff} - \frac{\gamma\alpha}{(1+\alpha^2)J_s}\vec{J}\times\left(\vec{J}\times\vec{H}_{eff}\right), \qquad (1)$$

where $\vec{J}$ is the magnetic polarization vector, $J_s$ is the magnetization polarization at saturation, $\gamma$ is the electron gyromagnetic ratio, $\alpha$ is the dumping parameter, and $\vec{H}_{eff}$ is the effective magnetic. The effective field, in general, is a superposition of four contributing fields: the externally applied field, the magnetocrystalline field, the demagnetizing field, and the exchange field, namely

$$\vec{H}_{eff} = \vec{H}_{ext} + \frac{2K_1}{J_s}\vec{n}\left(\vec{n}\cdot\frac{\vec{J}}{J_s}\right) + \vec{H}_{demag} + \frac{2A}{J_s}\Delta\left(\frac{\vec{J}}{J_s}\right), \qquad (2)$$

where $K_1$ is the magnetocrystalline anisotropy constant, $\vec{n}$ is the unit vector pointing in the magnetic easy-axis direction, and $A$ is the exchange constant.

## II. DISCUSSION OF HYSTERETIC RESULTS

The simulations were carried out with the MagPar[32] LLG micromagnetic solver using dynamic integration of the equation of motion and finite element meshing for precise sample shape visualization. Meshing was made from finite tetrahedral elements with dimensions of maximal 3.7 nm, which is smaller than the Py exchange length of 5.7 nm.[33] The meshing was approximately 10 times denser near the edges to include the influence of demagnetizing fields more strictly. For the other physical parameters we chose the exchange constant $A = 1.05 \cdot 10^{-11}$ J/m, the magnetic polarization at saturation $J_s = 1$ T, and the Gilbert damping constant $\alpha = 0.01$.[33]

Two different types of hysteresis loops have been examined. In the first type, the external magnetic field $H_{ext}$ has been applied along the x-axis (in the sample plane), in the second, the field has been applied along the z-axis (perpendicular to the base plane). In all cases, starting from a random state of magnetization at an external field $H_{ext} = 0$, the field has been changed with constant speed of 10 kA/(m·ns) up to 450 kA/m ($H_{ext}$ along x-axis) or to 600 kA/m ($H_{ext}$ along z-axis), respectively (this step is not shown in the graphs for clarity). Next, starting from the positive saturation state, the field has been swept at a constant speed of -10 kA/(m·ns) to -450 kA/m ($H_{ext}$ along x-axis) or to -600 kA/m ($H_{ext}$ along z-axis), respectively. In order to close the hysteresis loops, the field has been swept back from negative to positive saturation with the same constant speed afterwards. This field sweeping speed is comparable to values normally used in magneto-electronic applications.[34]

*2.1. Hysteresis loops for the in-plane applied field (x-direction)*

The hysteresis loops obtained for $H_{ext}$ applied along the x-axis are shown in Fig. 2. For the solid half-ball (Fig. 2a), an outstanding finding is the start of the magnetization reversal before $H_{ext} = 0$ is reached. Such a behaviour has also been observed by He *et al.* and interpreted as the beginning of a magnetic vortex state formation.[5] Our simulation strongly supports this interpretation, as can be seen in Fig. 2a where snapshots of the magnetization in z-direction are presented, showing vortex formations correlated with the start of the oscillations. A detailed analysis of the oscillation frequencies will be given in a future article.

While cutting the top of the solid half-ball suppresses magnetization reversal before $H_{ext} = 0$ (Fig. 2b), a hole in the sphere (Fig. 2c) significantly enhances the coercivity as well as the field region between first and second step of the complete magnetization reversal. This broad step is correlated with a flux-closed vortex state, while the saturation magnetization states and the oscillatory region are associated with two different onion states. The snapshots for the magnetization reversal from positive to negative saturation show the z-components of the magnetization. The oscillation here is not circular but a symmetric oscillation with the two coloured domain walls oscillating towards each other and back again. Following the snapshots of the x-component of the magnetization from negative to positive saturation, the magnetization reversal from the original onion state along a second onion state to the vortex state and finally to the reversed saturated onion state is clearly visible.

However, additional top cuts in a sphere with a hole (Fig. 2d) reduce the coercivity again, which is opposite to the finding in solid half-balls. Obviously, both effects have to be taken into account if a certain coercive field is desired.

*2.2. Hysteresis loops for the field applied perpendicular to a base-plane (z-direction)*

Figure 3 shows hysteresis loops for $H_{ext}$ applied along the z-axis. There are two types of reversal processes obtained, those with a step (Figs. 3a, 3b, and 3e) and those with smooth dynamics (Figs. 3c, 3d, and 3f). The insets in Fig. 3a show that in the solid half-ball a vortex core is built,[35] which is separated from the outer part of the vortex by a circular domain wall. During the magnetization reversal process, the core region becomes smaller, while the outer ring reverses. The oscillatory step finalizes the reversal by switching the core region. Both reversal processes from positive to negative saturation and vice versa differ by the orientation of the core and the outer region.

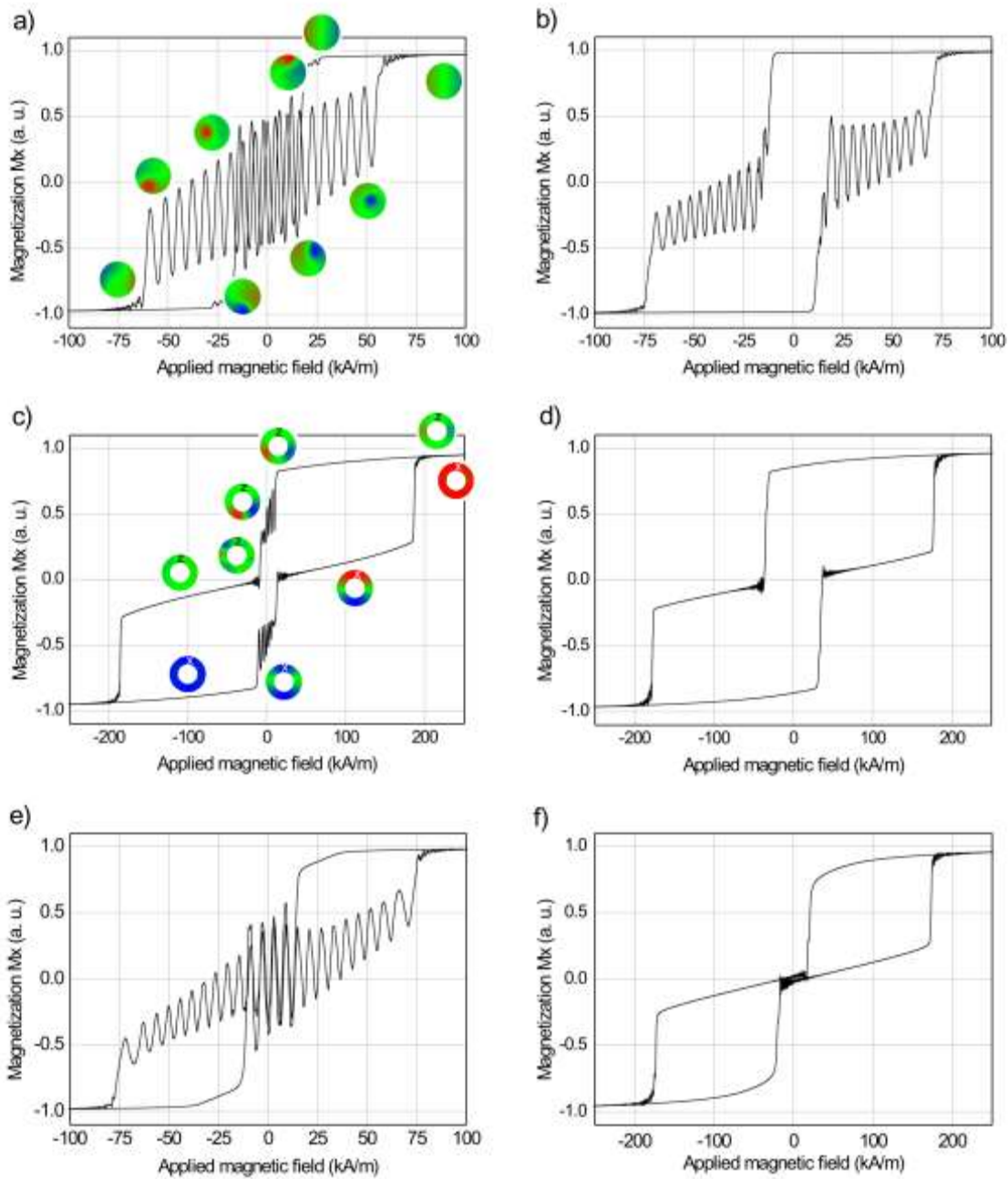

FIG. 2: Hysteresis loops, simulated for the samples depicted in figure 1, for the external field applied along the x-axis. The colour-coded insets in (a) show the z-component of the magnetization for different external fields, the insets in (c) exhibit the z-components (for field sweep from positive to negative values) and x-components (for field sweep from negative to positive saturation), respectively.

Apparently, the cylindrical cuts in the solid half-balls (Figs. 3b and 3e) lead to smaller coercivities and remanences, while the holes (Figs. 3c, 3d, and 3e) produce smooth reversible magnetization curves. Such a reversible magnetization curve of quasi-spherical Fe particles of 200 nm diameter has also been found by Diao et al.[36] However, investigation of the magnetization dynamics[35] and the colour-coded insets in Fig. 3c shows that here a flux-closed state is generated, corresponding to two possible states with flux rotational direction clockwise or counter-clockwise, enabling the utilization of these nano-magnets in storage devices as well. During the magnetization reversal process, the magnetization values as well as the colour-coded snapshots are identical for all external fields; the process is completely reversible.

Moreover, comparing Figs. 2 and 3, it is obvious that for the Py half-balls, the hard axis is always directed along the z-axis, as in 2D nano-magnets.

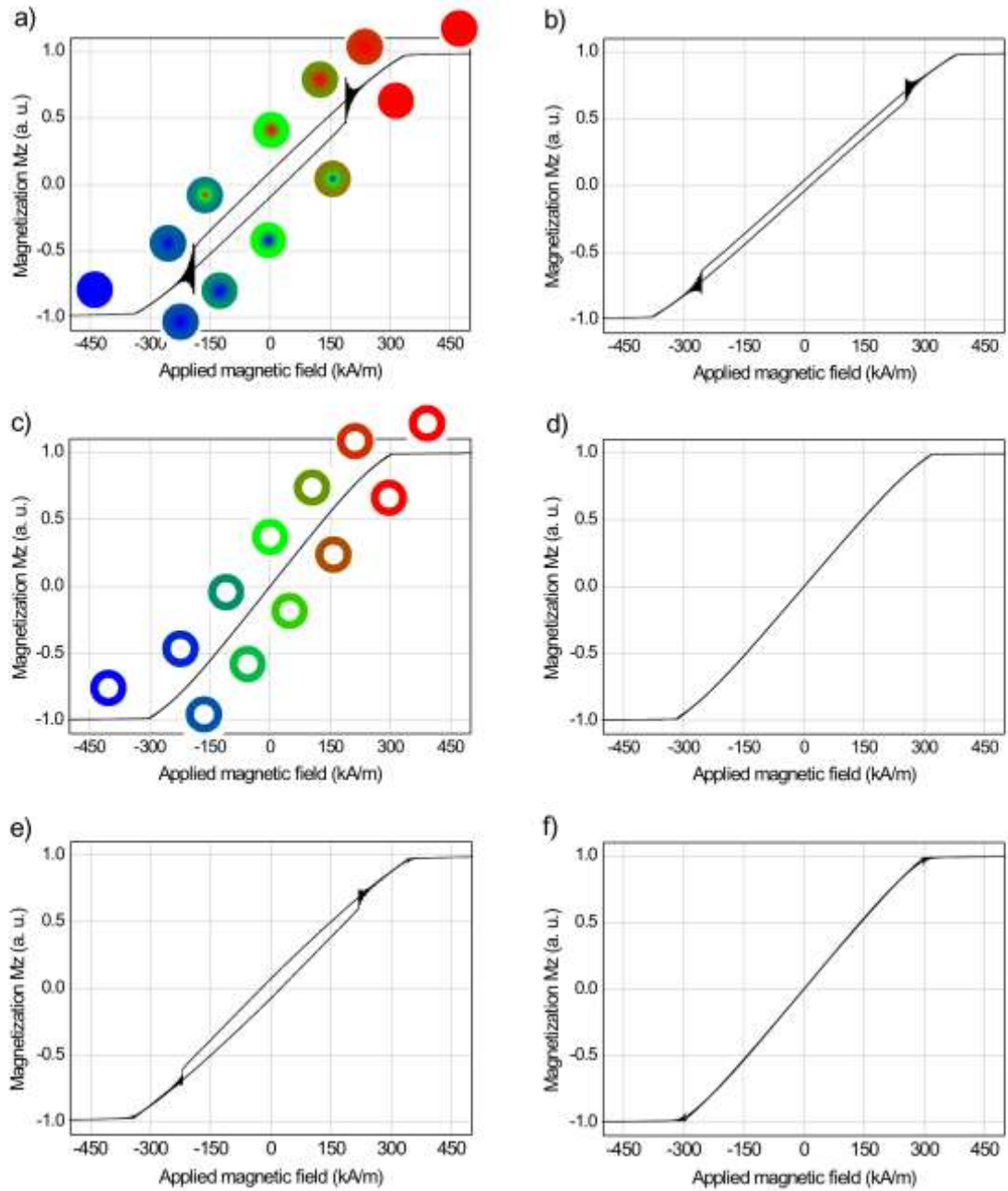

FIG. 3: Hysteresis loops, simulated for the samples depicted in figure 1, for the external field applied along the z-axis. The colour-coded insets in (a) and (c) show the z-component of the magnetization for different external fields.

## III. DISCUSSION OF MAGNETIC ENERGY DYNAMICS

In Figs. 4 and 5 demagnetizing and exchange energy evolutions in time are depicted for decreasing externally applied magnetic fields, equivalent to hysteretic evolutions from saturation to anti-saturation (compare Figs. 2-3). Since saturation and anti-saturation are states of the maximized demagnetizing and minimized exchange energies, a given magnetic sample evolves by intermediate states described by different scenarios. Generally, inside the intermediate regions exchange energy dominates due to occurrence of flux-closured states (Fig. 6). However, the demagnetizing fields, due to shape modifications, can radically change their spatial distribution influencing the exchange fields in an equivalent manner (Fig. 7). This influence affects the whole sample, volume and edges, due to the reduced dimensionality.

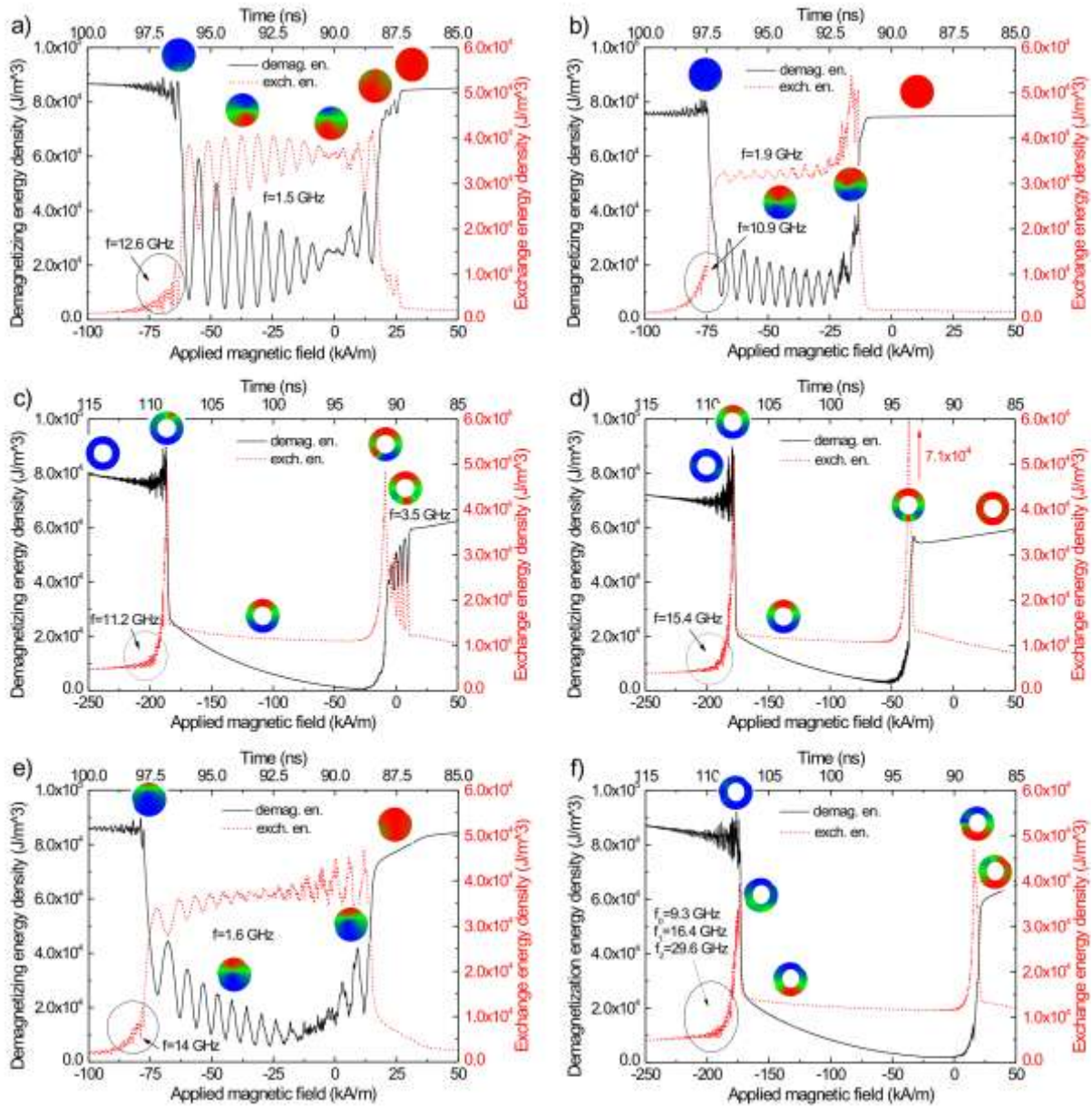

FIG. 4: Demagnetizing and exchange energy, simulated for the samples depicted in figure 1, for the external field swept along the x-axis from positive to negative saturation. The colour-coded insets show the x-component of the magnetization for different external fields.

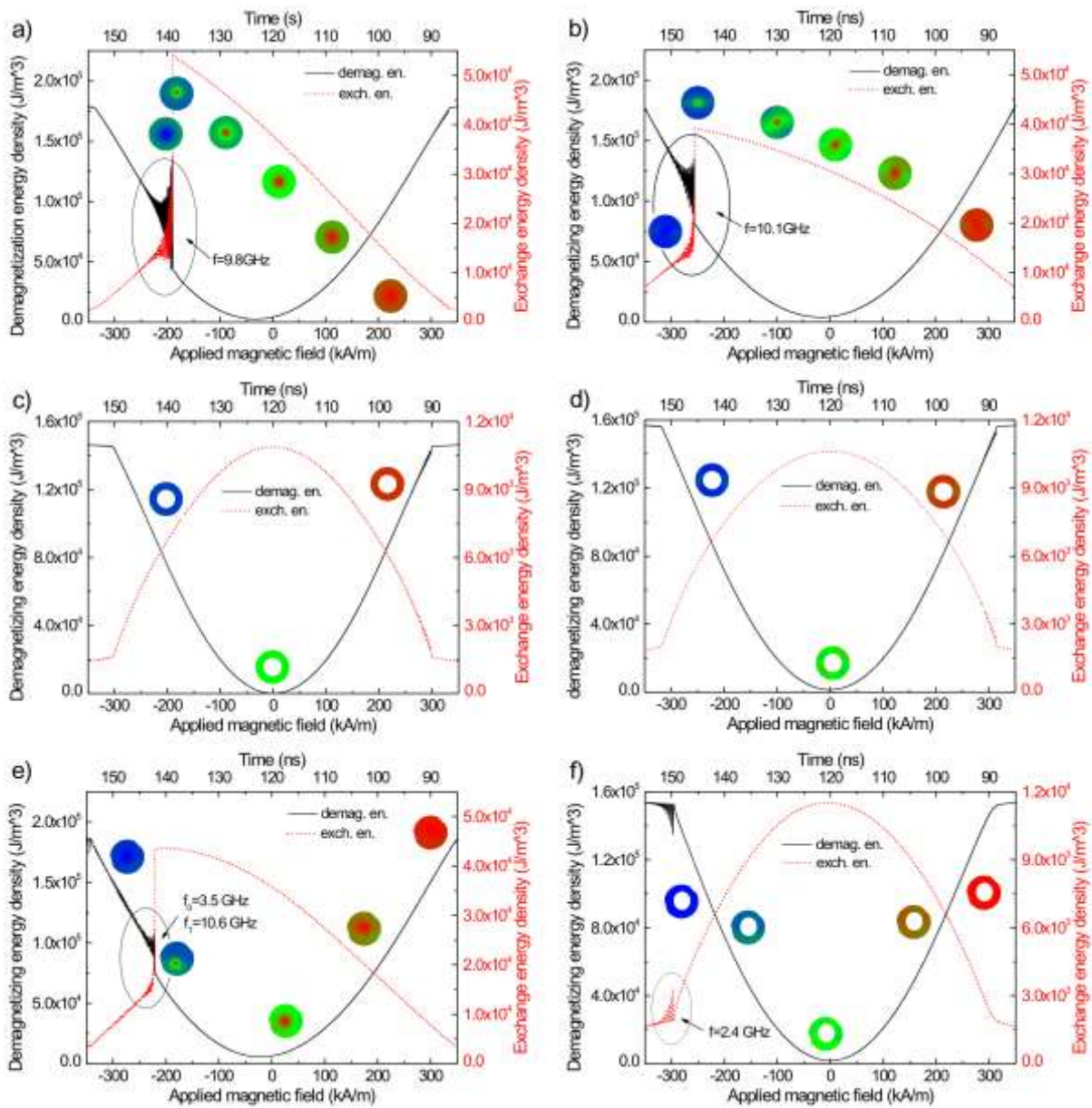

FIG. 5: Demagnetizing and exchange energy, simulated for the samples depicted in figure 1, for the external field swept along the z-axis from positive to negative saturation. The colour-coded insets show the z-component of the magnetization for different external fields.

*3.1. Energy dynamics for external magnetic field applied along x-axis*
Several regularities can be recognized in this case. Firstly, for samples without a hole (Figs. 1a, b, e), in the intermediate part of evolution, after passing the remanence, oscillations of the vortex core occur. The frequency of 1.5 GHz for a perfect solid sample (a) is lower than 1.9 GHz for the sample with a horizontal cut (b), and it is slightly increased to 1.6 GHz for the imperfect sample (e). The exchange and demagnetizing oscillations are anti-phased, however, amplitudes of these oscillations are relatively weaker for shape-modified samples (b) and (e). Importantly, the switching time is shorter for the sample with a cut (Fig. 4b) – where it is approximately 6 ns – than for the perfect solid sample (Fig. 4a), where it is equal to about 9.5 ns.

Secondly, for samples with a hole (Figs. 1c, d, f) in which the core of the vortex is excluded, the dynamics are different. The beginning of the intermediate-region evolution can happen via 3.5 GHz oscillations (Fig. 4c) or, for samples with cuts (Figs. 4d, f), these oscillations are excluded. These pre-intermediate oscillations are caused by formation of flux-closured states induced at the circumference edges – also for the solid case seen in Fig. 4a –; however, they are eliminated by magnetic poles located at intentionally introduced edges, also for samples without holes (Figs. 4b, e). While there are no energy oscillations in the intermediate regions for the drilled samples, however, there exists a rapid transient state for the exchange energy which can be attributed to the energy necessary to switch from the second onion state to the vortex state. The highest peak is observed for the sample with the horizontal cut (7.1 x $10^4$ J/m$^3$). Importantly, that peak is reduced for the imperfect sample (Fig. 4f), where the randomly oriented magnetic poles, located at the edges, act against this effect.

Thirdly, the intermediate regions are followed by relatively high-frequency oscillations in the 10-30 GHz range. Amplitudes of these oscillations are relatively lower than those of the lower frequency (below 2 GHz) intermediate-region oscillations. Generally, it seems that modifications of the shape cause an increase of these ripple frequencies, sometimes they even lead to occurrence of higher harmonics (Fig. 4f).

*3.2. Energy dynamics for external magnetic field applied along z-axis*

Energy dependencies for this case are much simpler than for case A. Since the direction of the externally applied magnetic field is parallel to the highest symmetry axis (z-axis), one observes regular, closed circulations of magnetization (Fig. 6). For samples with holes, energy evolutions are smooth and completely symmetrical (Figs. 5c, d). For the imperfect, deformed sample, as seen in Fig. 5f, 2.4 GHz oscillations occur which are followed by saturation. This behaviour is associated with competition between demagnetizing fields, located near the external circumference edge, and vortexes leaving a sample volume near that edge. In the three reversible cases for the nano-dots with holes and an external magnetic field applied along the z-axis (Figs. 5c, 5d, and 5f), both energies show roughly opposing values – this effect may be attributed to the lack of energy loss in a hysteretic loop.

For all remaining solid samples, asymmetrically located oscillations (Figs. 1a, b, e, f) can possess composed nature. Thus, the exchange energy maximum is phase-shifted in comparison to the demagnetizing energy maximum. Next, the following ripple oscillations bring back both energies to natural correlated, anti-phased behaviour, before the samples are saturated. This effect is shape-dependent, and the observed frequencies are in the range of 9.8-10 GHz. The reversal process is associated with competition between demagnetizing poles, located not only at sample circumference, but additionally at newly introduced edges. Surprisingly, the amplitude of these oscillations is smaller for the more deformed sample (Fig. 1e) in comparison to the more symmetrical cases (Figs. 1a, b).

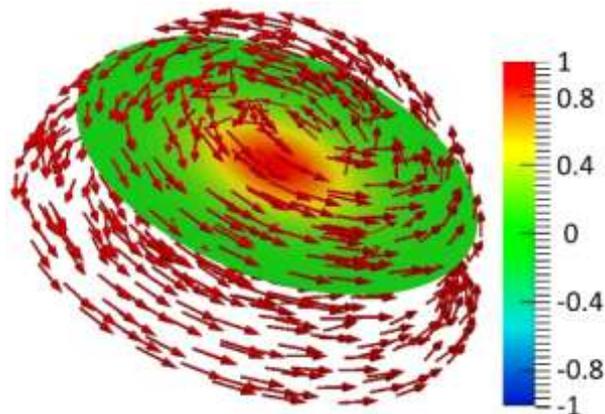

FIG. 6: 3-dimensional visualization of magnetization vectors for the external magnetic field applied along the half-ball symmetry axis (z-axis), for the solid half-ball (comp. figure 1a). Colours at the cross-section represent local values of $M_z$ magnetization components: 1 (red) - vector parallel to z-axis, 0 (green) - vector perpendicular to z-axis, -1 (blue) - vector antiparallel to z-axis. The figure represents zero-valued external-field state (remanence).

All the observed effects can be more deeply understood by considering Fig. 8, where both competing fields are visualized. It can be recognized that an additional horizontal cut increases the in-plane (x-y) components of the demagnetizing field blocking the exchange field precessions.

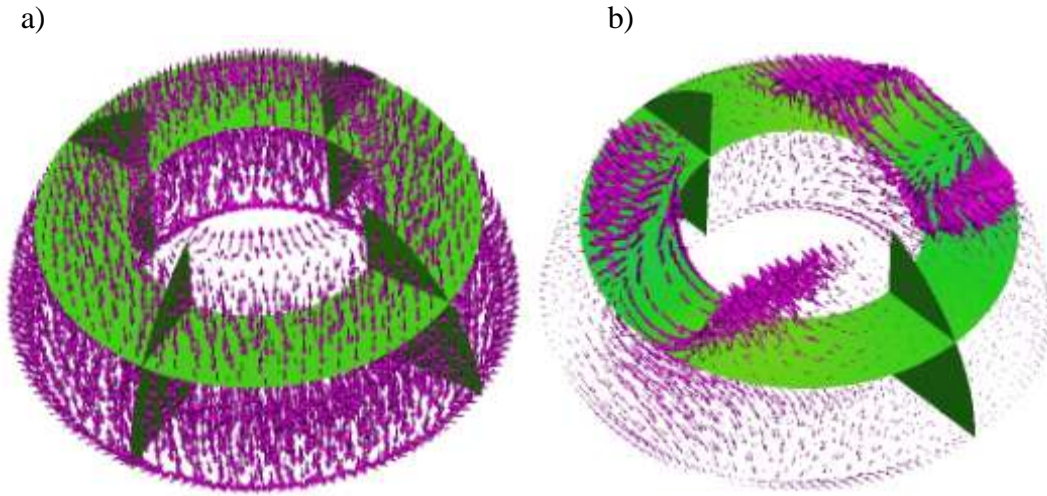

FIG. 7: 3-dimensional visualization of the demagnetizing field for the external magnetic field applied along the half-ball symmetry axis (z-axis, compare figure 1). In the solid half-ball with a hole (a), the field is concentrated at sample base circumference as well as at the hole circumference, while introduction of a horizontal cut completely changes the spatial distribution of the fields and the conditions for driving magnetization dynamics (b). The figure represents zero-valued external-field state (remanence).

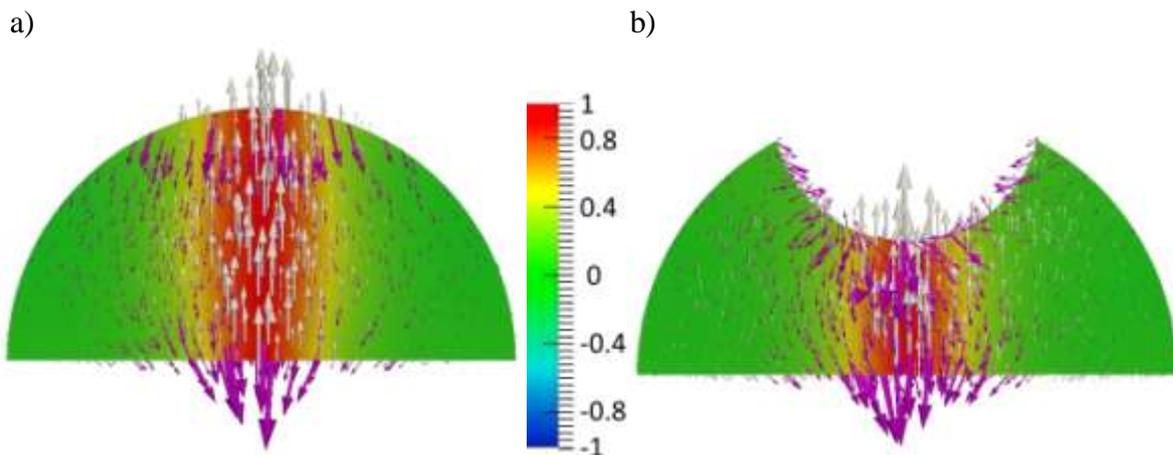

FIG. 8: Visualizations of cross-sections of competing demagnetizing (dark lilac arrows) and exchange fields (light grey arrows) for the full solid sample (a) and the sample with a single horizontal cut (b). The colours at cross-sections represent local values of the $M_z$ magnetization component: 1 (red) - vector parallel to z-axis, 0 (green) - vector perpendicular to z-axis, -1 (blue) - vector antiparallel to z-axis. The figure represents zero-valued external-field state (remanence).

## IV. CONCLUSION

In summary, we have provided some basic nanomagnetic sample 3D paradigms and given an overview about magnetization characteristics which can be reached by changes in the geometry of magnetic nano-particles, enabling deeper understanding of desired properties in a relatively broad range of modifications to meet the challenges of new 3D technology of magnetic devices. This could give an opportunity for new applications in magneto-electronics.